\documentclass[twocolumn,showpacs,preprintnumbers,amsmath,amssymb,prb,aps]{revtex4}

\usepackage{epsfig}
\usepackage{graphicx}
\usepackage{dcolumn}
\usepackage{bm}
\usepackage{ulem}

\begin{document}
\title{Evolution of the Kondo resonance feature and its relationship
to spin-orbit coupling across the quantum critical point in
Ce$_2$Rh$_{1-x}$Co$_x$Si$_3$}
\author{Swapnil Patil}
\author{V. R. R. Medicherla}
\author{R. S. Singh}
\author{E. V. Sampathkumaran}
\author{Kalobaran Maiti}
\altaffiliation{Corresponding author: kbmaiti@tifr.res.in}
\affiliation{Department of Condensed Matter Physics and Materials
Science, Tata Institute of Fundamental Research, Homi Bhabha Road,
Colaba, Mumbai 400 005, India}
\date{\today}
\begin{abstract}

We investigate the evolution of the electronic structure of
Ce$_2$Rh$_{1-x}$Co$_x$Si$_3$ as a function of $x$ employing high
resolution photoemission spectroscopy. Co substitution at the Rh
sites in antiferromagnetic Ce$_2$RhSi$_3$ leads to a transition from
an antiferromagnetic system to a Kondo system, Ce$_2$CoSi$_3$ via
the Quantum Critical Point (QCP). High resolution photoemission
spectra reveal distinct signature of the Kondo resonance feature
(KRF) and its spin orbit split component (SOC) in the whole
composition range indicating finite Kondo temperature scale at the
quantum critical point. We observe that the intensity ratio of the
Kondo resonance feature and its spin orbit split component, KRF/SOC
gradually increases with the decrease in temperature in the strong
hybridization limit. The scenario gets reversed if the Kondo
temperature becomes lower than the magnetic ordering temperature.
While finite Kondo temperature within the magnetically ordered phase
indicates applicability of the spin density wave picture at the
approach to QCP, the dominant temperature dependence of the
spin-orbit coupled feature suggests importance of spin-orbit
interactions in this regime.

\end{abstract}
\maketitle

\section{Introduction}

Quantum criticality \cite{Hertz} in heavy fermion systems is
obtained by tuning the antiferromagnetic phase transition
temperature to zero by external non-thermal control parameter like
pressure, chemical composition, magnetic field {\it etc}. The study
of quantum criticality has attracted a great deal of attention
recently due to the finding of non Fermi liquid excitations in the
properties of metals near the magnetic instability in correlated
systems and in particular in unconventional superconductors. The
quantum critical point (QCP) separates two entirely different ground
states - (i) long range magnetically ordered state and (ii)
paramagnetic/nonmagnetic Fermi liquid state.\cite{Doniach} While the
physics in the ground states on either sides, away from the QCP are
reasonably well understood, the physics in the vicinity of QCP is a
long standing puzzle.\cite{Si,Millis,Garnier} The question to be
answered is, how the quantum critical fluctuations affects the
electronic excitations and how it evolves as one goes from one side
of QCP to the other side.

\begin{figure}
 \vspace{-2ex}
\includegraphics [scale=0.4]{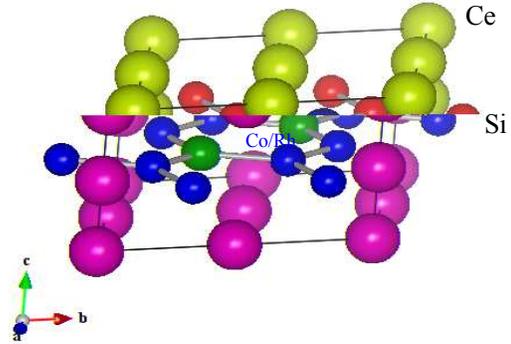}
\vspace{-40ex}
 \caption{Crystal structure of Ce$_2$Rh$_{1-x}$Rh$_x$Si$_3$ forming in
AlB$_2$ type hexagonal structure.}
\end{figure}

In order to address the issue, we studied an archetypical systems
Ce$_2$Rh$_{1-x}$Co$_x$Si$_3$ - the compounds in the whole
composition range crystallize in a AlB$_2$ derived hexagonal
structure with space group $P6/mmm$ as shown in Fig.
1.\cite{Patilbulk,Gordon,Majumdar} The valence band consists of Si
3$p$, Co 3$d$ and Rh 4$d$ electronic states.\cite{PatilRh,PatilCo}
Since the binding energy of the 4$d$ electronic states is higher
than the 3$d$ electronic states and 3$d$ electrons are relatively
more correlated due to smaller radial
extension,\cite{maiti3d,maiti4d} the substitution of Co at Rh sites
influences the effective $d$-$p$ hybridization as well as the
itineracy of the conduction electrons that leads to a plethora of
interesting properties. For example, one of the end members,
Ce$_2$RhSi$_3$ exhibit antiferromagnetic ordering at about 7
K.\cite{Patilbulk} Various studies involving lattice
compression/expansion via external pressure or chemical
substitutions suggested that this compound lies close to the peak of
Doniach's magnetic phase diagram.\cite{Nakano,Das} The sample with
$x$ = 0.6 composition exhibits quantum critical behavior such as
non-Fermi liquid behavior in the electrical transport measurements.
The compositions with $0.0 < x < 0.6$ exhibit antiferromagnetic
ordering. It was shown that application of high pressure on the
compounds in this composition range leads to a change in properties
similar to that observed via change in composition.\cite{jpcs11} The
magnetic susceptibility measurements on the other end member,
Ce$_2$CoSi$_3$, does not show magnetic ordering down to 0.5 K akin
to a Kondo lattice compound with an estimated Kondo temperature,
$T_K$ of about 44 K. The temperature dependence of electrical
resistivity of Co-end members appear to be similar to a typical
mixed valent Kondo lattice system.\cite{Patilbulk,lawrence}

In this article, we present our results of investigation on the
evolution of electronic structure as a function of Rh/Co
concentration employing high resolution photoemission spectroscopy.
The high resolution spectra exhibit distinct signatures of Kondo
resonance feature and its spin orbit coupled component in the whole
composition range. The relative intensity of the Kondo resonance
feature with respect to the spin orbit split feature exhibits
distinctly different temperature evolution in the different parts of
the phase diagram.

\section{Experimental details}

High quality samples of Ce$_2$Rh$_{1-x}$Co$_x$Si$_3$ for various
values of $x$ were prepared in the polycrystalline form by arc
melting - growth of the material from congruently molten phase
ensured large grain size with homogenous stoichiometry. The samples
are characterized by x-ray diffraction, transport and magnetic
measurements as described elsewhere.\cite{Patilbulk} The
photoemission measurements were carried out using high resolution
Gammadata Scienta electron analyzer, SES2002 and monochromatic
photon sources at a base pressure better than 3 $\times$ 10$^{-11}$
torr. The energy resolutions were set to 300 meV at  Al $K\alpha$
(1486.6 eV) photon energy, and 5 meV at He {\scriptsize II} (40.8
eV) and He {\scriptsize I} (21.2 eV) photon energies. The samples
were very hard and could not be cleaved or fractured. Hence, the
sample surface was cleaned by {\it in situ} scraping at each
temperature using a fine grained diamond file and the surface
cleanliness was ensured by the absence of O 1$s$ and C 1$s$ signals.
The reproducibility of the spectra was ensured after each trial of
surface cleaning. The Fermi level was determined at each temperature
by measuring the Fermi cutoff in the spectral function of high
purity silver mounted in electrical contact with the sample. The
temperature variation down to 5 K was achieved using a open cycle
helium cryostat, LT-3M from Advanced Research Systems, USA.

\section{Results and Discussions}

\begin{figure}
 \vspace{-2ex}
\includegraphics [scale=0.4]{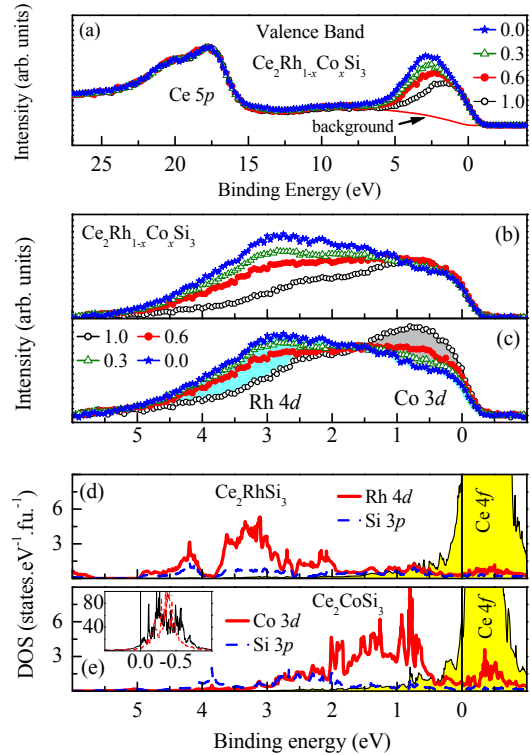}
\vspace{-4ex}
 \caption{(a) XP valence band spectra of Ce$_2$Rh$_{1-x}$Co$_x$Si$_3$
for various values of $x$ at 5 K. The solid line shows a typical
integral background function. (b) Background subtracted high
resolution valence band spectra exhibiting evolution of the spectra
with $x$ at 5 K. (c) The same background subtracted spectra shown
after normalization at 1.5 eV binding energy to demonstrate how the
spectral intensity changes with composition. The gray shade shows
the increase in Co 3$d$ intensity and blue shade shows subsequent
decrease in Rh 4$d$ intensities with the increase in $x$. LDA
results of the partial density of states of (d) Ce$_2$RhSi$_3$ and
(e) Ce$_2$CoSi$_3$. The $y$-scale is expanded to visualize the Ce
4$f$ (yellow shaded region), Si $p$ and Co/Rh $d$ partial density of
states (PDOS) in the occupied part with clarity. The inset in (e)
shows the Ce 4$f$ PDOS for Ce$_2$CoSi$_3$ (dashed line) and
Ce$_2$RhSi$_3$ (solid line).}
\end{figure}

The photoemission measurements were carried out on four
compositions, namely, $x$ = 0.0 (antiferromagnetic), 0.3 (30\% Co
substituted and antiferromagnetic), 0.6 (corresponds to QCP), 1.0
(Kondo compound) that spans the whole range of the solid solution
with varied properties. In Fig. 2(a), we show $x$-ray photoemission
(XP; $h\nu$ = 1486.6 eV) valence band spectra collected at 5 K for
the above four compositions (energy resolution used for these
spectra = 0.6 eV). The spectra primarily consist of two groups of
features. The signals corresponding to Ce 5$p_{3/2}$ and 5$p_{1/2}$
photoemission appear at 17.5 eV and 20.3 eV binding energies,
respectively. The features close to the Fermi energy, $\epsilon_F$
constitute the valence band. Since Ce concentration is same in all
the samples, the spectra have been normalized by the Ce 5$p$ peak
intensity. It is evident that the spectral weight around 2.8 eV
gradually reduces with the decrease in Rh content.

To probe this region with greater clarity, we have obtained the high
resolution (resolution = 0.3 eV) XP spectra of the valence band
region close to $\epsilon_F$ as shown in Fig. 2(b) after integral
background subtraction - a typical background function is shown in
Fig. 2(a) by solid line. The spectral intensity within 1 eV of
$\epsilon_F$ does not change significantly with $x$ - a small
increase is observed around 0.5 eV while there is a large decrease
in intensity around 2.8 eV with the increase in Co content. Such an
anomalous change can be attributed to the large photoemission
cross-section of Rh 4$d$ states ($\sim$ 0.012) compared that for Co
3$d$ states ($\sim$ 0.0037).\cite{Yeh} In order to visualize the
spectral evolution due to substitution better, we renormalized the
same spectra by the intensity at 1.5 eV binding energy as shown in
Fig. 2(c). The increase in intensity corresponding to Co 3$d$
photoemission is shown by gray shade and the decrease Rh 4$d$
intensity is shown by blue shade - the gradual change in spectral
intensities with the increase in $x$ is beautifully manifested in
the figure. These results suggest dominant Co 3$d$ contributions
close to $\epsilon_F$, whereas Rh 4$d$ contribute at higher binding
energies.

The energy band structure calculations were performed for the end
members Ce$_2$RhSi$_3$ and Ce$_2$CoSi$_3$ using full potential
linearized augmented plane wave method (FLAPW) within the local
density approximations using Wien2k software.\cite{wien} The lattice
constants and all other structural details were obtained from the
literature.\cite{Patilbulk} The calculated Ce 5$p$, Rh 4$d$/Co 3$d$
and Si 3$p$ partial density of states (PDOS) for Ce$_2$RhSi$_3$ and
Ce$_2$CoSi$_3$ are shown in Fig. 2(d) and Fig. 2(e), respectively.
The results reveal dominance of the contributions of Rh/Co $d$
states in the valence band. The Rh 4$d$ PDOS appear around 3 eV
binding energies consistent with the experimental observations. The
Co 3$d$ states peak around 1 eV binding energy. Clearly, the
calculated results correspond well with the experimental spectra
shown in Fig. 2(c).

The $d$ PDOS seem to have influenced the conduction electron-Ce 4$f$
hybridization ($cf$-hybridization) significantly - Ce 4$f$ PDOS
width is broader in Ce$_2$RhSi$_3$ relative to that in
Ce$_2$CoSi$_3$ as shown in the inset of Fig. 2(e). Moreover, the
intensity of Co 3$d$ PDOS in the Ce 4$f$ region is stronger than the
Rh 4$d$ contributions in the same energy range. These observations
suggest stronger Co 3$d$-Ce 4$f$ hybridization than Rh 4$d$-Ce 4$f$
hybridization. Evidently, proximity of the $d$ states have strong
influence on $cf$-hybridization\cite{Gunnarsson1} in addition to the
influence of lattice volume change often referred in these
systems.\cite{Patilbulk}

\begin{figure}
 \vspace{-2ex}
\includegraphics [scale=0.4]{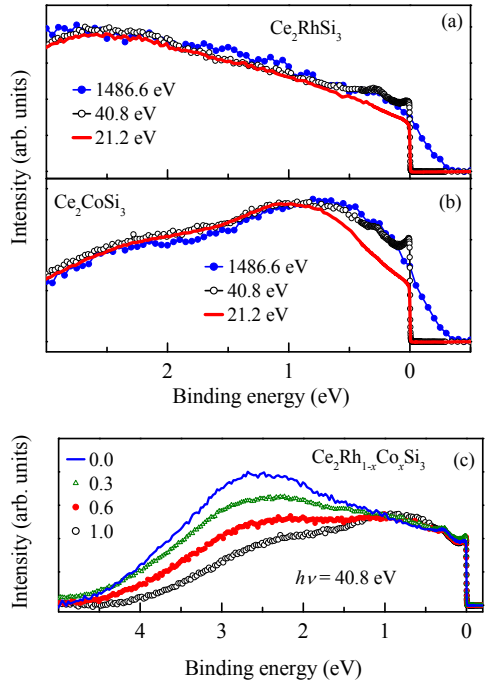}
\vspace{-8ex}
 \caption{Valence band spectra of (a) Ce$_2$RhSi$_3$ and (b)
Ce$_2$CoSi$_3$ at 5 K obtained using different photon energies
exhibiting results consistent with the band structure results. (c)
He {\scriptsize II} spectra of Ce$_2$Rh$_{1-x}$Co$_x$Si$_3$ at 5 K
for various values of $x$. The high energy resolution employed in
this technique reveal distinct Kondo features near the Fermi level.}
\end{figure}

In Fig. 3, we compare the photoemission spectra collected using He
{\scriptsize I}$\alpha$, He {\scriptsize II}$\alpha$ and Al
$K\alpha$ photon energies at 5 K. All the spectra are normalized by
the intensities between 1 - 2.5 eV binding energy range. The
features in He {\scriptsize II} and XP spectra are quite similar
except the fact that the XP spectra are broader due to higher
resolution broadening. Two distinct and sharp features appear in the
vicinity of $\epsilon_F$ in the He {\scriptsize II} spectra of both
Ce$_2$RhSi$_3$ and Ce$_2$CoSi$_3$.\cite{PatilRh,PatilCo} The He
{\scriptsize I} spectra, however, do not exhibit these intensities.
Considering contributions of various electronic states near
$\epsilon_F$ depicted in Figs. 2(d) and 2(e), and the dependence of
their photoemission cross section,\cite{Yeh} the sharp features near
the Fermi level can be attributed primarily to the photoemission
from Ce 4$f$ electronic states. The feature near $\epsilon_F$
corresponds to 4$f_{5/2}$ photoemission and is a manifestation of
the Kondo resonance features. The corresponding spin-orbit split
component, the 4$f_{7/2}$ final state appears at a relatively higher
binding energies ($\sim$~260 meV). In Fig. 3(c), we show the He
{\scriptsize II} spectra of all the compositions at 5 K. Although
the Rh 4$d$ intensity gradually decreases with the increase in $x$
as also observed in the XP spectra, the features near $\epsilon_F$
seems similar in the whole composition range.

Considering significant enhancement of the Ce 4$f$ contributions in
the He {\scriptsize II} spectra relative to the other contributions
in the same energy region, one can extract the Ce 4$f$ spectral
function by subtracting the He {\scriptsize I} spectra from the He
{\scriptsize II} spectra as often done in similar Ce-based systems.
The peak of the Kondo resonance feature usually appears just above
the Fermi energy (unoccupied) and the photoemission spectroscopy
probes the higher binding energy tail of the Kondo
peak.\cite{Patthey,Ehm} Thus, the extracted 4$f$ spectral weight by
the difference (He {\scriptsize II} - He {\scriptsize I}) depicting
the occupied part of the Ce 4$f$ PDOS would provide a good
estimation of the Kondo resonance feature.

\begin{figure}
 \vspace{-2ex}
\includegraphics [scale=0.4]{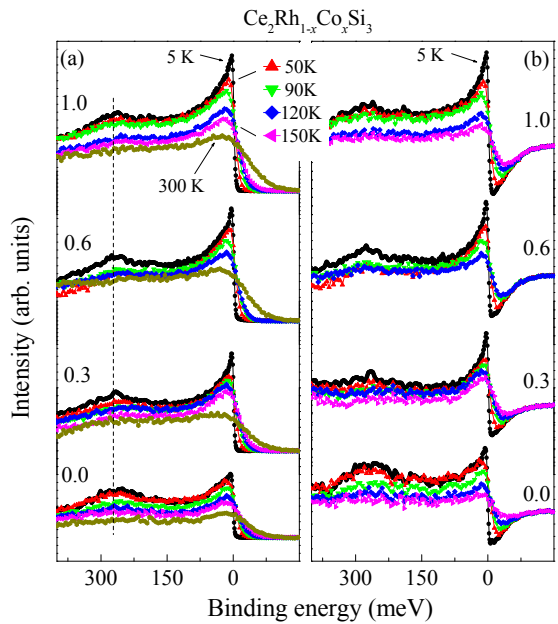}
\vspace{-12ex}
 \caption{(a) The evolution of the extracted Ce 4$f$ spectral functions
with temperature for different compositions. (b) The spectral weight
transfer obtained by subtracting the room temperature spectra from
the spectra at different temperatures.}
\end{figure}

The (He {\scriptsize II} - He {\scriptsize I}) spectra at different
temperatures are shown in Fig. 4(a) for different compositions
separately. The spectra of all the compositions exhibit a sharp
feature near $\epsilon_F$ - the intensity of the feature reduces
gradually with the increase in temperature suggesting presence of
Kondo compensation effect in every case despite the fact that the
magnetism in these materials are widely varied. The intensity of the
Kondo resonance feature at a particular temperature gradually
increases with the increase in $x$. Since the Kondo temperature
increases with the increase in $x$, the above observation suggests
that the intensity of the Kondo resonance feature is directly
related to the Kondo temperature, $T_K$.\cite{Garnier} While such a
scenario is expected in a Kondo system, the similar observation in
long range ordered systems ($x <$ 0.6) indicates the existence of a
finite Kondo temperature scale even in these cases.

The Kondo coherence temperature for the intermediate compositions
estimated\cite{Patilbulk} from the electrical transport properties
is $\sim$ 8 K for $x$ = 0.3 and $\sim$ 50 K for $x$ = 0.6. When the
measurement temperature is below the coherence temperature, there
will be additional complications due to the expansion of the Fermi
volume and presumably manifested in the evolution of the Kondo
resonance feature. The Kondo temperatures for the compositions, $x$
= 0.0, 0.3 and 0.6 are 12 K, 30 K and 40 K,
respectively.\cite{Patilbulk} The photoemission spectra collected at
120 K and 150 K are well above the coherence temperature, Kondo
temperature and N\'{e}el temperature for all the compositions and
still exhibit finite intensity of the Kondo resonance feature. At
all the temperatures, we find qualitatively similar trend of the
evolution of the Kondo resonance feature across the solid solution
indicating that the spectral weight redistributions, if any, due to
the onset of the Kondo coherence are not altering the inferences
drawn. This manifests applicability of the single impurity Kondo
physics in the electronic structure\cite{Klein} and the Kondo
temperature scale remains finite at the QCP.

The intensity of the corresponding spin-orbit split feature
(4$f_{7/2}$ signal) appearing around 260 meV binding energy (marked
by a vertical dashed line in Fig. 4(a)) in each of the samples also
reduces as the temperature is increased. Interestingly, the
intensity of the 4$f_{7/2}$ signal seems to be similar in all the
compositions although the intensity of the Kondo resonance feature
at the Fermi level increases with the increase in Co concentration.
At finite temperature, the spectral intensity representing the
occupied part of the density of states becomes lower at $\epsilon_F$
due to the thermal effect represented by the Fermi-Dirac
distribution function. The Fermi-Dirac distribution function
depletes the intensity below $\epsilon_F$ and enhances the intensity
above $\epsilon_F$ in a symmetric manner with the increase in
temperature. We have subtracted the room temperature spectra from
the spectra at different temperatures ($I(\epsilon,T) -
I(\epsilon,300K)$) to investigate the spectral weight transfer with
temperature. Such exercise would provide a change in spectral weight
symmetric with respect to the Fermi level in a Fermi liquid system.
The subtracted spectral functions, shown in Fig. 4(b), however,
exhibit asymmetric spectral evolutions with temperature - the gain
of spectral weight below the Fermi level is larger than the loss of
weight above the Fermi level. Such an effect appear in the presence
of a peak such as Kondo resonance peak in the vicinity of the Fermi
level. The asymmetry seems to be weaker towards Rh end indicating
weaker Kondo peak. This clearly indicates that the temperature
dependence of the Kondo peak and the spin-orbit peak is different.

\begin{figure}
 \vspace{-2ex}
\includegraphics [scale=0.4]{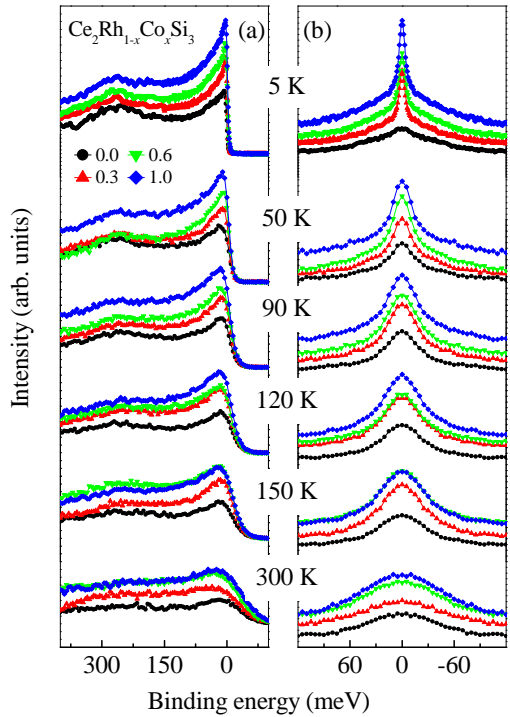}
\vspace{-2ex}
 \caption{(a) The evolution of the extracted Ce 4$f$ spectral functions
with composition, $x$ at different temperatures. (b) The spectral
functions obtained by symmetrizing the experimental spectra to
estimate the spectral weight the Fermi level.}
\end{figure}

To investigate the spectral changes with composition, $x$ at a
particular temperature, we re-plot the Ce 4$f$ spectral functions as
a function of composition at a particular temperature in Fig. 5(a).
The experimental spectra exhibit gradual increase in intensity of
both the peaks with the increase in $x$ that can be associated to
the change in the Kondo temperature as discussed above and also
observed earlier in other Ce-based compounds.\cite{Garnier}

We compare the evolution of the intensities of the Kondo resonance
peak and its spin-orbit split component in Fig. 6. In order to
deconvolve the Fermi Dirac distribution induced depletion of the
spectral density of states at the Fermi level, we symmetrized the
spectra with respect to the Fermi level - if $I(\epsilon)$
represents the spectral function, the symmetrized spectra can be
expressed as $I(\epsilon) = I(\epsilon - \epsilon_F) + I(\epsilon_F
- \epsilon)$. It is often found that the intensity at the Fermi
level can be estimated quite accurately by such symmetrization
process, which is independent of the presence/absence of
particle-hole symmetry. The symmetrized spectra are shown in Fig.
5(b) those represent the intensity of the Kondo resonance feature
(KRF).

\begin{figure}
 \vspace{-2ex}
\includegraphics [scale=0.4]{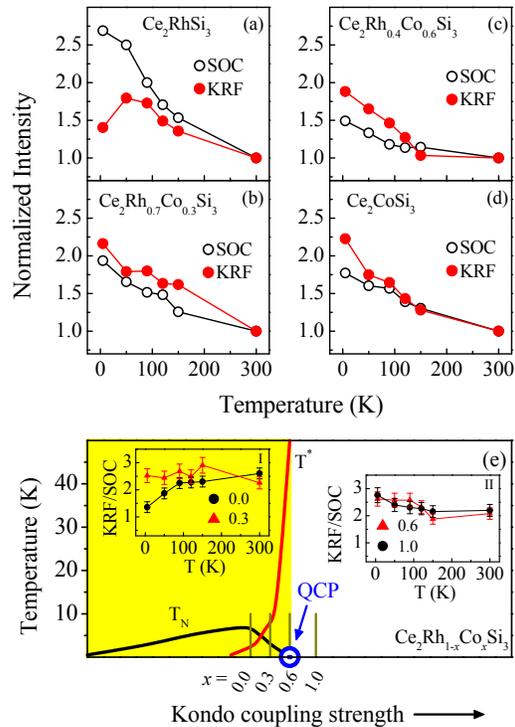}
\vspace{-4ex}
 \caption{Intensity variation of the Kondo resonance
feature (KRF) and its spin orbit coupled (SOC) component as a
function of temperature of (a) Ce$_2$RhSi$_3$, (b)
Ce$_2$Rh$_{0.7}$Co$_{0.3}$Si$_3$, (c)
Ce$_2$Rh$_{0.4}$Co$_{0.6}$Si$_3$ and (d) Ce$_2$CoSi$_3$. (e) A
schematic phase diagram showing the long range ordered region
(yellow) and Kondo compensated region separated by a quantum
critical point (QCP). The left and right insets show the intensity
ratio of KRF/SOC as a function of temperature for corresponding
compositions in Ce$_2$Rh$_{1-x}$Co$_x$Si$_3$. $T_N$ and $T^*$
represent the Ne\'{e}l temperature and Kondo coherence temperature
and are obtained from Ref. [6].}
\end{figure}

The estimated intensities of the spin orbit peak (SOC) obtained from
Fig. 4(a) and the Kondo peak (KRF) obtained from Fig. 5(b) are shown
in Fig. 6(a) (for Ce$_2$RhSi$_3$), 6(b) (for
Ce$_2$Rh$_{0.7}$Co$_{0.3}$Si$_3$, 6(c) (for
Ce$_2$Rh$_{0.4}$Co$_{0.6}$Si$_3$ and 6(d) (for Ce$_2$CoSi$_3$).
Interestingly, the compositions exhibiting long range order show
weaker temperature dependence of the Kondo resonance feature (KRF)
compared to the spin orbit coupled (SOC) peak - the trend gets
reversed on the other side of the phase diagram. This is shown by
plotting the relative intensity ratio (KRF/SOC) in the insets of
Fig. 6(e) in the corresponding phase regime. In the weak
hybridization limit, the intensity ratio, KRF/SOC decreases with the
decrease in temperature - the Kondo peak grows less rapidly than the
spin orbit peak. On the other side (strong coupling limit), the
ratio, KRF/SOC gradually increases with the decrease in temperature.
Interestingly, the trend at the quantum critical point follows the
strong coupling behavior.

Kondo effect essentially arises from the hybridization of the
conduction electrons with the impurity moment. It is well known that
in the strong coupling limit, the electrons corresponding to the
local moment forms a quantum mechanically entangled states with the
conduction electrons, termed as Kondo singlet. In such a scenario,
the Fermi surface expands due to the additional contribution from
the electrons forming the local moment - here the 4$f$ electrons.
The Kondo temperature essentially provides an energy scale deriving
the expansion of the Fermi volume.\cite{SIFujimori,gegenwart1} In
many cases, it is observed that the Kondo temperature scale vanishes
at the approach to QCP.\cite{Klein,gegenwart2} The results in the
present study reveal finite Kondo temperature scale at the QCP in
the solid solution Ce$_2$Rh$_{1-x}$Co$_x$Si$_3$ suggesting that the
magnetic Ce 4$f$ electrons near QCP acquire itineracy at lower
temperatures - the antiferromagnetism near the QCP occurs among the
itinerant electrons and hence indicates the applicability of an SDW
picture in these cases.\cite{PatilRh,jpcs11} Interestingly, the
results in Fig. 6(e) suggests that if the Kondo temperature is
higher than N\'{e}el temperature, the intensity of the Kondo
resonance feature exhibits stronger temperature dependence compared
to its spin-orbit split counterpart. The scenario gets reversed, if
the Kondo temperature is lower than the ordering temperature.

\section{Conclusions}

In summary, we studied the evolution of the electronic structure of
Ce$_2$Rh$_{1-x}$Co$_x$Si$_3$ using high resolution photoemission
spectroscopy. The signature of Kondo resonance feature is revealed
in the high resolution low temperature spectra of all the compounds
providing evidence of finite Kondo temperature scale at the quantum
critical point. Interestingly, the temperature dependence of Kondo
resonance feature is found to be stronger than its spin-orbit split
counterpart if the Kondo temperature is higher than the magnetic
ordering temperature. The spin-orbit part show stronger temperature
dependence if the Kondo temperature becomes lower than the ordering
temperature. While finite Kondo temperature scale in this regime
establishes applicability of the spin density wave picture at the
approach to the quantum criticality, the spin-orbit coupling appears
to play a significant role in this regime. These results provide an
insight to the origin of quantum fluctuations often observed in the
proximity of the onset of exotic phases like superconductivity.

\section{Acknowledgements}

One of the authors S.P., thanks the Council of Scientific and
Industrial Research, Government of India for financial support.

\end{document}